\documentclass[aps,twocolumn,nofootinbib,superscriptaddress,amsfonts,floatfix
]{revtex4-1} 
\usepackage{graphicx,amsmath,amssymb,amstext}
\usepackage{amssymb,amsbsy,amsfonts,amsthm,color}

\usepackage{mathtools,nccmath}

\usepackage{epsfig}
\usepackage{graphicx}
\usepackage{subfigure}
\usepackage[dvipsnames]{xcolor}
\definecolor{linkcolor}{rgb}{0.0,0.3,0.5}
\usepackage[unicode, colorlinks=true, linkcolor=linkcolor, citecolor=linkcolor, 
filecolor=linkcolor,urlcolor=linkcolor, pdfusetitle]{hyperref}
\usepackage{cancel}
\usepackage{balance}
\usepackage[capitalise]{cleveref}
\usepackage{xspace}
\usepackage{enumerate}
\usepackage{ulem}
\usepackage{mdframed}
\usepackage{booktabs}
\usepackage{makecell}
\usepackage{enumitem,tabularx,multirow}
\usepackage{aas_macros}
\usepackage{colortbl}
\usepackage{orcidlink}

\AtBeginDocument{%
  \heavyrulewidth=.08em
  \lightrulewidth=.05em
  \cmidrulewidth=.03em
  \belowrulesep=.65ex
  \belowbottomsep=0pt
  \aboverulesep=.4ex
  \abovetopsep=0pt
  \cmidrulesep=\doublerulesep
  \cmidrulekern=.5em
  \defaultaddspace=.5em
}

\graphicspath{{Figures/}}

\usepackage{color}

\begin{document}

\title{Characterizing the  quark-hadron mixed phase in compact star cores : sensitivity to nuclear saturation and quark-model parameters at finite-temperature}

\author{Suman Pal,\orcidlink{0009-0000-5944-4261}}
\email{sumanvecc@gmail.com}
\affiliation{Physics Group, Variable Energy Cyclotron Centre, 1/AF Bidhan Nagar, Kolkata 700064, India}
\affiliation{Homi Bhabha National Institute, Training School Complex, Anushakti Nagar, Mumbai 400085, India}

\author{Gargi Chaudhuri,\orcidlink{0000-0002-8913-0658}}
\email{gargi@vecc.gov.in}
\affiliation{Physics Group, Variable Energy Cyclotron Centre, 1/AF Bidhan Nagar, Kolkata 700064, India}
\affiliation{Homi Bhabha National Institute, Training School Complex, Anushakti Nagar, Mumbai 400085, India}

\begin{abstract}

A thorough knowledge of the quark–hadron phase transition in hot and dense matter is essential for
constraining the equation of state of neutron stars.
In this work, we study the thermodynamics of the quark–hadron mixed phase at finite
temperature using the Gibbs construction and examine its impact on hybrid  star
matter. We systematically explore the role of nuclear saturation properties, including the
effective nucleon mass $m^*/m$, incompressibility $K_{\text{sat}}$, symmetry energy 
coefficient $J$, and its slope $L$, together with quark matter parameters such as the bag
constant $B_0^{1/4}$ and the vector coupling strength $G_V$.
We find that the width of the mixed phase is mainly controlled by the effective mass and
symmetry energy, while the roles of incompressibility and symmetry energy slope are
comparatively weak, particularly at higher temperatures.
Thermal effects substantially modify the phase structure: increasing temperature
reduces the mixed-phase width and softens the equation of state in the coexistence region
due to Gibbs phase equilibrium constraints.
These effects are reflected in the behavior of the speed of sound, the trace anomaly,
and its derivative.
Variations in the symmetry energy, effective mass, and quark parameters significantly affect the hadron–quark transition, stellar radii, and maximum mass, while finite temperature softens the equation of state and enhances radius jumps in the mixed phase. Strong vector repulsion is essential to reconcile massive pulsar observations with NICER constraints, whereas weaker repulsion favors more compact, low-mass configurations

\end{abstract}

\maketitle


\section{Introduction}

Compact stars serve as exceptional astrophysical laboratories for probing the properties of dense nuclear matter\cite{2022NatAs}. Extensive observations of pulsars \cite{Fonseca:2021wxt,Riley:2019yda,Miller:2019cac,Miller:2021qha}, combined with the groundbreaking detection of gravitational waves \cite{LIGOScientific:2017vwq,LIGOScientific:2018cki} have significantly advanced our understanding and  have 
placed important constraints on the nuclear equation of state (EoS). However, despite these advancements, the EoS still has substantial uncertainties, posing a persistent challenge in accurately modeling the internal structure of compact stars.

The phenomenon of hadron-quark phase transition is one of the most interesting current topics of research in compact star physics. 
In order to model the transition between hadronic and quark phases in neutron star interiors, researchers commonly employ either the Gibbs or the Maxwell approach, depending on the value of the surface tension at the hadron-quark interface \cite{Maruyama:2007ss,Maruyama:2007ey}.
A crucial physical constraint in the modeling of compact stars is charge neutrality \cite{Glendenning:1992vb}. Since gravitational binding dominates the stability of compact stars, the presence of any significant net charge would reduce this binding, making charge neutrality an essential global condition. However, within the mixed phase of hadronic and quark matter, this condition may be satisfied either locally, by demanding that each phase is separately charge neutral, or globally, by requiring only the net charge of the system to vanish. While both approaches satisfy the overall charge neutrality condition, they lead to different internal structures and energetics. 
Moreover, the global charge neutrality approach is consistent with the Gibbs conditions for phase equilibrium, which demand equality of temperature, chemical potentials, and pressure between the coexisting phases. In this work, we adopt the global charge neutrality condition in modeling the mixed phase, ensuring consistency with both the thermodynamic equilibrium and the physical conditions prevailing in the interior of compact stars.

In the context of hybrid stars, the choice between Gibbs \cite{Orsaria:2013hna,Nicotra:2006eg,Burgio:2007eg,Bhattacharyya:2009fg,Pagliara:2009dg,Hempel:2009vp,Yasutake:2009sh,Wu:2017xaz,Steiner:2000bi,Pagliara_2009,Hempel:2009vp,Yasutake:2009sh,Glendenning:1992vb,Glendenning:2001pe,Orsaria:2013hna,Wu:2018zoe,Han:2019bub,Rather:2020lsg,Ju:2021hoy,Ju:2021nev,Karimi:2022kcx,Liu:2023gmq,Liu:2023kzy,Mariani:2023kdu,Steiner:2000bi,Menezes:2003xa,Menezes:2003pa,Santos:2004js,Burgio:2007eg,Pagliara_2009} and Maxwell constructions \cite{Pal:2023quk,Pal:2025skz} becomes particularly important, since a small surface tension at the hadron–quark interface allows the Gibbs construction, combined with global charge neutrality, to provide a physically consistent description of the mixed phase.

In this work, we investigate the EoS of warm, beta-equilibrated neutrino free hadronic and quark matter, with a particular focus on its implications for the internal composition of compact stars. Specifically, we explore the properties of the mixed phase comprising both hadronic and quark matter and examine the potential existence of a pure quark matter core within such stars. Our analysis is carried out at  finite temperature, corresponding respectively to conditions relevant for proto-neutron stars and the remnants formed in binary neutron star mergers.
For the pure hadronic phase, we employ the well-known Relativistic Mean Field (RMF) model. For the quark matter phase, we consider the MIT bag model with vector interactions.

The primary motivation of this work is two-fold, spanning both microscopic thermodynamic modeling and macroscopic astrophysical implications.
From a theoretical perspective, we perform a detailed investigation of the hadron–quark mixed phase, focusing on the width of the mixed phase and its sensitivity to fundamental nuclear-matter and quark-matter parameters. In particular, we examine how key nuclear properties—such as nuclear effective mass, incompressibility and symmetry-energy coefficients and quark-matter parameters such as the bag constant and vector interaction strength—govern the onset, extent, and termination of the phase transition. We further explore the impact of finite temperature on the EoS, emphasizing its role in modifying the phase boundaries.

From an astrophysical perspective, these micro-physical features are directly imprinted on the global properties of neutron stars. Since the mass–radius relation of a neutron star is uniquely determined by the underlying EoS, understanding how microscopic parameters control the mixed-phase width is essential for interpreting present and future observations. 
Measuring neutron-star radii is very difficult, although their masses can often be measured accurately. The Neutron Star Interior Composition Explorer (NICER), mounted on the International Space Station, helps  to address this challenge by observing soft thermal X-ray emission from rotation-powered millisecond pulsars (MSPs). This emission comes from heated polar caps on the neutron-star surface and can be modeled to estimate the star’s mass and radius. Observations of X-ray emission from neutron-star surfaces therefore provide an important way to study their internal structure and constrain the equation of state of dense matter.  To date, NICER has enabled mass and radius inferences for five pulsars \cite{Hoogkamer:2025eaq}: PSR J0030+0451 with mass 1.44$_{-0.14}^{+0.15}M_{\odot}$ and radius 13.02$_{-1.06}^{+1.24}$ km, PSR J0437{\textendash}4715 \cite{Choudhury:2024xbk} with canonical mass 1.418$ \pm 0.037M_{\odot}$ and radius 11.36 $_{-0.63}^{+0.95}$ km, PSR J0740+6620 with mass 2.0073$\pm 0.069 M_{\odot}$ and radius 12.49 $_{-0.88}^{+1.28}$ km , PSR J1231-1411 \cite{Salmi:2024bss,Qi:2025mpn} with low mass 1.04$_{-0.03}^{+0.05}M_{\odot}$ and radius 13.5 $_{-0.5}^{+0.3}$ km, and PSR J0614-3329 \cite{Mauviard:2025dmd} with nearly canonical mass 1.44$_{-0.07}^{+0.06}M_{\odot}$ and radius 10.29 $_{-0.86}^{+0.1.01}$ km.
At the opposite end of the mass spectrum, the central compact object in the supernova remnant HESS J1731–347 \cite{Doroshenko:2022nwp} has been inferred to have a remarkably low mass, $M = 0.77^{+0.20}_{-0.17},M{\odot}$, and a small radius, $R = 10.4^{+0.86}_{-0.78}$ km. This source represents the lightest neutron star known to date and has been suggested as a possible candidate for an exotic compact object, such as a strange quark star. Such observations strongly motivate a careful examination of phase transitions in dense matter and their influence on stellar structure across a wide mass range.
In addition to X-ray observations, gravitational-wave detections by LIGO–Virgo–KAGRA provide complementary constraints on the neutron star EoS through measurements of tidal deformability \cite{LIGOScientific:2018cki,LIGOScientific:2020zkf}. These multimessenger observations collectively demand a unified framework in which thermodynamic properties of dense matter are consistently linked to astrophysical observables.

Therefore, by systematically connecting the thermodynamic characteristics of the hadron–quark mixed phase to neutron-star mass–radius relations and observational constraints, this work aims to bridge microphysical modeling with current and emerging astrophysical data, offering deeper insight into the nature of ultra-dense matter.

\section{Formalism}

\subsection{Hadronic matter}
We employ the RMF model with the BigApple parameterization \cite{Fattoyev:2020cws} to describe
the hadronic matter, where nucleons interact through the exchange of various mesons
including the isoscalar--scalar $\sigma$ meson, the isoscalar--vector $\omega$ meson,
and the isovector--vector $\rho$ meson.
The Lagrangian density for hadronic matter consisting of nucleons ($p$ and $n$) and
leptons ($e$) is written as
\cite{Menezes:2003xa,Menezes:2003pa,Pagliara_2009,Santos:2004js,Burgio:2007eg}
{\small
\begin{eqnarray}
\label{eq:LRMF}
\mathcal{L}&=&\bar{\psi}(i\gamma_{\mu}\partial^{\mu}-m_N)\psi+\frac{1}{2}(\partial_{\mu}\sigma\partial^{\mu}\sigma-m_{\sigma}^2\sigma^2)-\frac{1}{4}\omega_{\mu\nu}\omega^{\mu\nu} \nonumber\\
&+&\frac{1}{2}m_{\omega}^2\omega_{\mu}\omega^{\mu}-\frac{1}{4}\rho_{\mu\nu}\rho^{\mu\nu}+\frac{1}{2}m_{\rho}^2\vec{\rho_{\mu}}\vec{\rho^{\mu}}
 \nonumber\\
 &+& (g_{\sigma}\bar{\psi}\sigma\psi -g_{\omega}\bar{\psi}\gamma_{\mu}\omega^{\mu}\psi -\frac{1}{2} g_{\rho}\bar{\psi}\gamma_{\mu}\vec{\tau}.\vec{\rho^{\mu}})-\frac{1}{3}bm(g_{\sigma}\sigma)^3       \nonumber\\
 &-&\frac{c}{4} (g_{\sigma}\sigma)^4  +\Lambda_{\omega}g_{\omega}^2(\omega_{\mu}\omega^{\mu})(g_{\rho}^2\vec\rho_{\mu}\vec{\rho^{\mu}})+\frac{\zeta}{4!}(g_{\omega}^{2}\omega_{\mu}
      \omega^{\mu})^{2}\nonumber\\
 &+&\sum_{l= e}\bar{\psi_l}(i\gamma_{\mu}\partial^{\mu}-m_l)\psi_l  ~.
\end{eqnarray} }

The mean field equations are :

\begin{eqnarray} \label{mean_field_eq}
   m_\sigma^2 \,  \sigma_0&=& -m \, b \,  g_{\sigma}^3 \sigma_0^2 - c 
     g_{\sigma}^4  \sigma_0^3 + g_{\sigma} ( \rho_p^s+\rho_n^s) , \nonumber \\  
  m_\omega^2 \, \omega_0 &= &- \frac{\zeta}{3!}  g_{\omega}^4 \omega_0^3
     - 2 \Lambda_{\omega} g_{\rho}^2  g_{\omega}^2  \rho_{03}^2  \omega_0 +
     g_{\omega} \, (\rho_p+\rho_n) , \nonumber \\  
  m_\rho^2 \,   \rho_{03} &=& - 2 \Lambda_{\omega} g_{\rho}^2  g_{\omega}^2
     \omega_0^2 \rho_{03}+  \frac{g_{\rho}}{2} \,(\rho_p-\rho_n) \ .  
\end{eqnarray}

where the scalar $\rho_i^s$ and vector $\rho_i$ 
densities at finite temperature are given by
\begin{eqnarray}
\label{eq:bary_dens}
\rho_i&=& = \frac{\gamma_i}{2 \pi^2}\int_0^{\infty}  dk \, k^2 \, (f_{i}(k,T) - f_{\bar{i}}(k,T))  , \nonumber \\
\rho_i^s&=& \frac{\gamma_i}{2 \pi^2}\int_0^{\infty}  dk \, k^2 
\, \frac{m^*_i}{\sqrt{k^2+m_i^{*2}}}\left(f_{i}(k,T)+f_{\bar{i}}(k,T)\right),~~
\end{eqnarray}
with $\gamma_i=2$ accounting for the degeneracy of the spin degree of freedom of baryons and 
\begin{equation}
f_{i/\bar{i}}(k,T) =\left[1+\text{exp}\left(\frac{\sqrt{k^2+m_i^{*2}} \mp \mu_i^{*}}{T}\right)\right]^{-1},
\label{eq:distribution}
\end{equation}

\begin{equation}
\mu_i^{*} = \mu_i - g_{\omega} \omega_{o}  -  \tau_{3i}\frac{g_{\rho}}{2}\rho_{03}
\label{eq:mueff}
\end{equation}
where $\tau_{3i}$  being the appropriate isospin projector ($\tau_{3p}=1$ and $\tau_{3n}=-1$).
The expression of the energy density, pressure and entropy density for the hadronic natter is given by, 
\begin{eqnarray}
\label{eq:energy-pressure}
\epsilon_{H} &=&\frac{1}{2\pi^2}\sum_{i} \gamma_i \int_0^{\infty} dk k^2\sqrt{k^2 + m_i^{*2}}\,(f_{i}(k,T) + f_{\bar{i}}(k,T)) \nonumber \\
&& +  \frac{1}{2\pi^2} \sum_{l} \gamma_l \int_0^{\infty} dk k^2\sqrt{k^2 + m_l^{2}}\, (f_{l}(k,T) + f_{\bar{l}}(k,T)) \nonumber \\
&& + \frac{1}{2}(m_{\omega}^2 \omega_0^2+m_{\rho}^2 \rho_{03}^2+ m_{\sigma}^2\sigma_0^2)+\frac{b}{3}(g_{\sigma } \sigma_0)^3  +\frac{c}{4}(g_{\sigma } \sigma_0)^4 \nonumber \\
&& + \frac{\zeta}{8}(g_{\omega }\omega_0)^4 + 3\Lambda_{\omega}(g_{\rho }g_{\omega } \rho_{03} \omega_0)^2 ,\nonumber \\
P_{H} &=& \frac{1}{6\pi^2}\sum_{i} \gamma_i \int_0^{\infty} dk \frac{k^4}{\sqrt{k^2 + m_i^{*2}}}{(f_{i}(k,T) + f_{\bar{i}}(k,T))}  \nonumber \\
&&+ \frac{1}{6\pi^2}\sum_{l} \gamma_l \int_0^{\infty} dk \frac{k^4}{\sqrt{k^2 + m_l^{2}}}(f_{b}(k,T) + f_{\bar{b}}(k,T))  \nonumber \\
&& + \frac{1}{2}(m_{\omega}^2 \omega_0^2+m_{\rho}^2 \rho_{03}^2- m_{\sigma}^2\sigma_0^2)-\frac{b}{3}(g_{\sigma } \sigma_0)^3  -\frac{c}{4}(g_{\sigma } \sigma_0)^4 \nonumber \\
&& + \frac{\zeta}{24}(g_{\omega }\omega_0)^4 + \Lambda_{\omega}(g_{\rho }g_{\omega } \rho_{03} \omega_0)^2 ,\nonumber \\
&&  s_{H} = \frac{1}{T}\left(P_H+\varepsilon_H -\sum_{i}\mu_i \rho_i\right).  \nonumber \\
\end{eqnarray}

The saturation properties at $\rho_0$, namely the binding energy per nucleon ($B/A$), 
nuclear incompressibility ($K_{\text{sat}}$), symmetry energy coefficient ($J$), 
and symmetry energy slope parameter ($L$), are listed in Table~\ref{tab:1}.
To explore the influence of bulk nuclear matter properties on the mass and radius 
of neutron stars, a mapping between the coupling constants of the interaction 
Lagrangian and the empirical nuclear matter properties must be established. 
In this work, we follow the calibration procedure described in Ref.~\cite{Chen:2014sca}.
The coupling constants $g_{\sigma}$, $g_{\omega}$, $b$, and $c$ are obtained 
by reproducing the empirical values of $B/A$, $K_{\text{sat}}$, and the 
$m^*$ at $\rho_0$. 
The remaining two couplings, $g_{\rho}$ and $\Lambda_{\omega}$, 
are determined as a function
of the symmetry energy $J$ and its slope parameter $L$. 
Further details of the calibration procedure can be found in Ref.~\cite{Chen:2014sca,Hornick:2018kfi}.
In this work, we investigate the sensitivity to four nuclear matter parameters 
by varying them individually around the BigApple EoS saturation point. 
When one parameter is varied, the remaining three are held fixed 
at their original saturation values of the BigApple model.

\begin{table*}[!ht]
\caption{The nuclear matter properties at saturation density $\rho_0$for the BigApple EoS \cite{Fattoyev:2020cws}.  }
\setlength{\tabcolsep}{20.0pt}
\begin{tabular}{cccccc}
\hline
\hline
 $\rho_{0}$ & $B/A$ & $K_{\text{sat}}$ &$m^*/m$  & $J$ &$L$\\
 $({\rm fm}^{-3})$ & (MeV) & (MeV) &  &(MeV) & (MeV)\\ \hline
 0.155 & $-$16.344 & 227.001 & 0.608& 31.315 & 39.80 \\
\hline
\hline
\end{tabular}
\label{tab:1}
\end{table*}

\color{black}

\begin{figure*}[htbp] 
    \centering
    \includegraphics[width=0.32\textwidth]{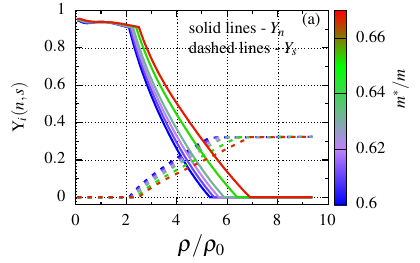}
    \includegraphics[width=0.32\textwidth]{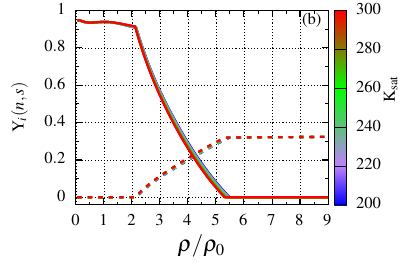}
    \includegraphics[width=0.32\textwidth]{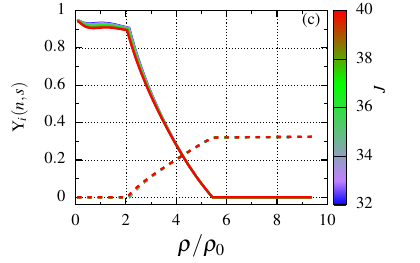}
    \includegraphics[width=0.32\textwidth]{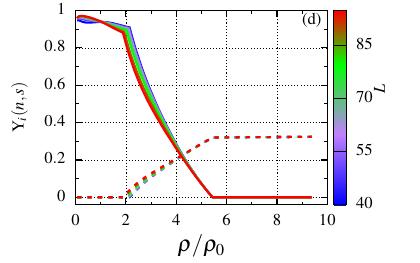}
    \includegraphics[width=0.32\textwidth]{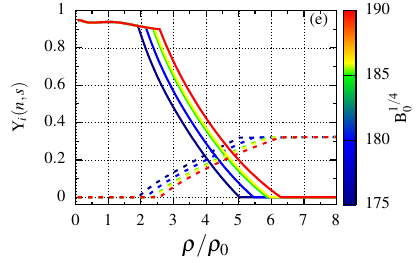}
    \includegraphics[width=0.32\textwidth]{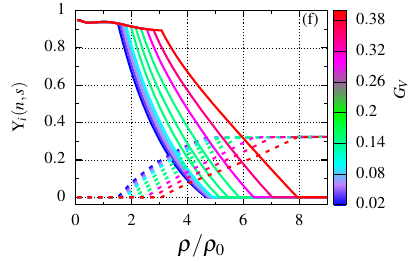}
    \caption{  Composition of hybrid-star matter: variation of neutron and strangeness fractions with baryon density (in units of the saturation density,) for different hadronic and quark model parameters.     }
    \label{fig:part_frac}
\end{figure*}

\subsection{Quark matter}
The description of quark matter is described by MIT bag model with vector interaction \cite{Pal:2023dlv,Pal:2024nza}. 
The thermodynamic potential  can be  then written as \cite{Pal:2023dlv,Pal:2024nza,Dexheimer:2013eua},
\begin{equation}\label{eq:quasi_pressure}
\begin{aligned}
    \Omega_Q=&-\sum_{i=u,d,s,}\frac{1}{3}\frac{\gamma_i}{2\pi^2}\int_0^{\infty}\frac{k^4}{\sqrt{k^2+(m_i^*)^2}}[f_i^++f_i^{-}] dk-\\ &\sum_{l=e} \frac{1}{3}\frac{\gamma_l}{2\pi^2}\int_0^{\infty}\frac{k^4}{\sqrt{k^2+m_l^2}}[f_{l}^{+}+f_l^{-}]+B_0 -\frac{1}{2}m_V^2V_0^2 \\
\end{aligned}
\end{equation} 
The quark degeneracy factors are $\gamma_{u}=\gamma_{d}=\gamma_{s}=6$ and for electron $\gamma_e=2$.
The Fermi distribution functions for particle and antiparticle are : $f_i^\pm= \frac{1}{1+exp\left({\frac{E_i(k)\mp \mu_i}{T}}\right)} $ , where $E_i(k)=\sqrt{k^2+(m_i)^2}$.\\
The inclusion of a vector meson in the description of quark matter leads to modifications in the equation of state. The effect of the vector interaction is taken through the following relation $\mu_i=\mu_i^*+g_{\text{V}}\text{V}_0$. 
Here $g_{V}$ is the vector interaction coupling and $m_V$ is mass of vector mesons.
The equation of motion for the vector field is given by :
\begin{align}
\frac{\partial \Omega} {\partial V_0}=0\implies m_V^2 V_0 = \sum_{i=u,d,s} g_{V} \rho_i 
\end{align} 
The vector interaction coupling constant is expressed as $G_V=\left(\frac{g_V}{m_V}\right)^2$.
The quark number densities are given by 
\begin{equation} \label{eq:quasi_density}
\begin{aligned}
     \rho_i&=\frac{\gamma_i}{2\pi^2}\int_0^{\infty}k^2[f_i^{+}-f_i^{-}] dk\\  
\end{aligned}
\end{equation}
The expression of pressure is given by,
\begin{equation}
    P_Q=-\Omega_Q
\end{equation}
The expression of the energy density is given by,
\begin{equation} \label{eq:quasi_energy_density}
\begin{aligned}
  \varepsilon_Q = & \sum_{i=u,d,s}  \frac{\gamma_i}{2\pi^2} \int_0^{\infty} k^2 \sqrt{k^2 + (m_i^*)^2} \left( f_i^{+} + f_i^{-} \right) dk+ B_0 \\
   &  +  \frac{\gamma_e}{2\pi^2}\int_0^{\infty} k^2 \sqrt{k^2 + m_e^2} \left( f_{e}^{+} + f_e^{-} \right) dk + \frac{1}{2}m_V^2V_0^2\\
\end{aligned}
\end{equation}

In this work, the quark matter phase is modeled using the MIT bag model with a repulsive vector interaction, providing a simple framework to systematically study the hadron–quark phase transition. While quark superfluidity (color superconductivity \cite{Kurkela:2024xfh,Sen:2022qol,Agrawal:2009ad}) could affect the stiffness of the quark equation of state at high densities, its inclusion would introduce additional parameters ( pairing gaps $\Delta$). We therefore focus on the dominant effects arising from the quark sector mainly the role of the strong repulsive interaction and bag constant.

\subsection{Mixed Phase}
In the mixed phase\cite{Glendenning:1992vb,Glendenning:2001pe,Orsaria:2013hna,Wu:2018zoe,Han:2019bub,Rather:2020lsg,Ju:2021hoy,Ju:2021nev,Karimi:2022kcx,Liu:2023gmq,Liu:2023kzy,Mariani:2023kdu,Steiner:2000bi,Menezes:2003xa,Menezes:2003pa,Santos:2004js,Burgio:2007eg,Pagliara_2009}, the charge neutrality condition, Gibbs phase equilibrium conditions, and relevant thermodynamic quantities are defined as follows:

\medskip
\noindent
Charge Neutrality Condition:
\begin{equation}
(1 - \chi)~ \rho_c^{H} + \chi~\rho_c^{Q} = 0~,
\end{equation}
where $\chi$ is the volume fraction of the quark phase, and $\rho_c^{H}$, $\rho_c^{Q}$ are the charge densities in the hadronic and quark phases, respectively.

\medskip
\noindent
Thermodynamic Quantities in the Mixed Phase:

The total energy density, baryon density, and entropy per baryon in the mixed phase are obtained as:
\begin{equation}
\varepsilon = (1 - \chi)\, \varepsilon_{H} + \chi\, \varepsilon_{Q}  ~,
\end{equation}
\begin{equation}
\rho = (1 - \chi)~ \rho_{H} + \chi~ \rho_{Q} ~,
\end{equation}
\begin{equation}
\frac{s}{\rho} = (1 - \chi)~ \left( \frac{s}{\rho} \right)_{H} + \chi~ \left( \frac{s}{\rho} \right)_{Q} ~.
\end{equation}

\section{Results }
\label{sec:results}

\subsection{Thermodynamic behavior of EOS}

We aim to investigate the effects of nuclear saturation parameters, namely
$m^*/m$, $K_{\text{sat}}$, $J$, and $L$, as well as quark matter parameters,
namely $G_V$ and $B_0^{1/4}$, on the properties of the mixed phase associated
with the quark--hadron phase transition in the core of neutron stars.
In addition, we analyze the influence of temperature on the structure and
extent of the mixed-phase region.

\textbf{\textit{Particle fraction :}}
To characterize the composition of the mixed phase, we focus on two representative particle fractions: the neutron fraction
$Y_n = \rho_n / \rho_B$ from the hadronic sector, and  the strangeness fraction
$Y_S = \rho_s / (3\rho_B)$ from the quark sector.
Figure~\ref{fig:part_frac}(a) shows the variation of these particle fractions with
$m^*/m$.
We find that, as $m^*/m$ increases, the neutron abundance in the mixed phase increases.
 The Gibbs phase equilibrium conditions require a redistribution of
the quark degrees of freedom, which suppresses the strange quark population in the mixed phase with increasing
$m^*/m$.
Figure~\ref{fig:part_frac}(b) illustrates the dependence  on
 nuclear incompressibility $K_{\text{sat}}$ and it is observed that 
 
 variation in $K_{\text{sat}}$ has  a minor effect on both the
neutron and strangeness fractions throughout the mixed phase.

Figures~\ref{fig:part_frac}(c) and \ref{fig:part_frac}(d) show the variation of the
particle fractions with the symmetry energy parameters, namely the symmetry energy
coefficient $J$ and its slope $L$.
We find that  both these parameters primarily affect the hadronic phase  and the early stage of mixed phase by altering the isospin composition of dense matter;  the effect of $L$  extends to higher densities as compared to $J$ in the mixed phase. This behavior suggests that, in the density range relevant for the mixed phase, the effects of the symmetry energy—are strongly suppressed by the Gibbs phase equilibrium conditions and the increasing contribution
of quark matter.
\begin{figure}[htbp] 
    \centering
    \includegraphics[width=0.50\textwidth]{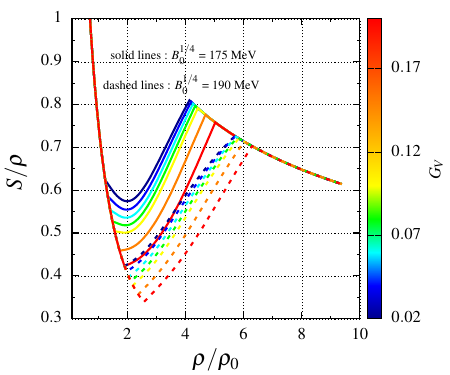}
    \caption{ Entropy per baryon as a function of baryon density (in units of the saturation density) for different strengths of the repulsive quark interaction in hybrid stars.  }
    \label{fig:entropy}
\end{figure}

\begin{figure*}[htbp] 
    \centering
    \includegraphics[width=0.45\textwidth]{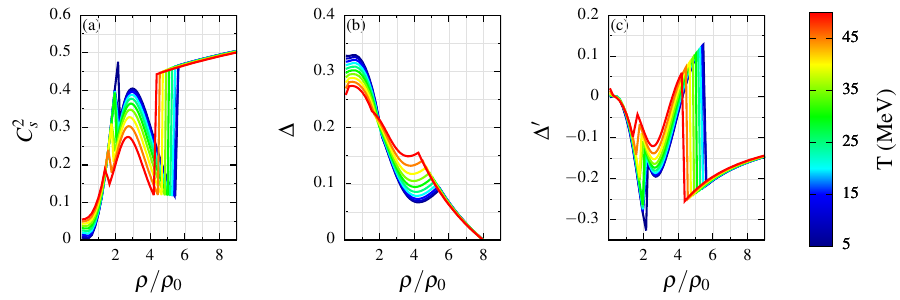}
    \includegraphics[width=0.45\textwidth]{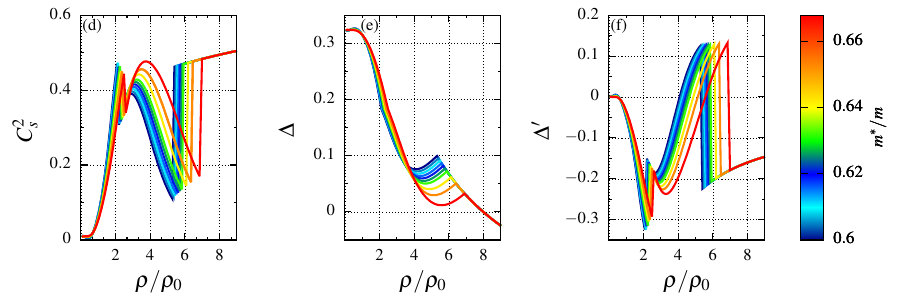}
    \includegraphics[width=0.45\textwidth]{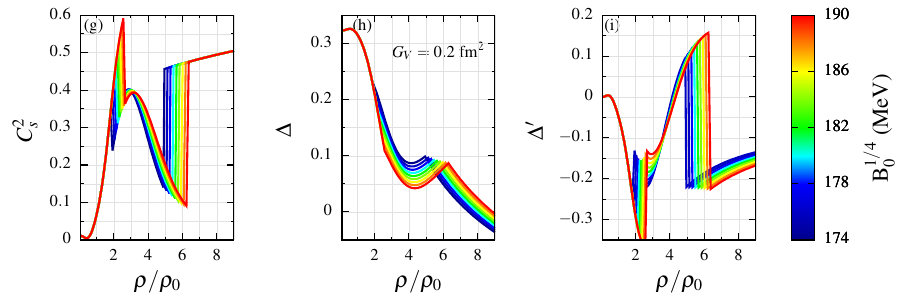}
    \includegraphics[width=0.45\textwidth]{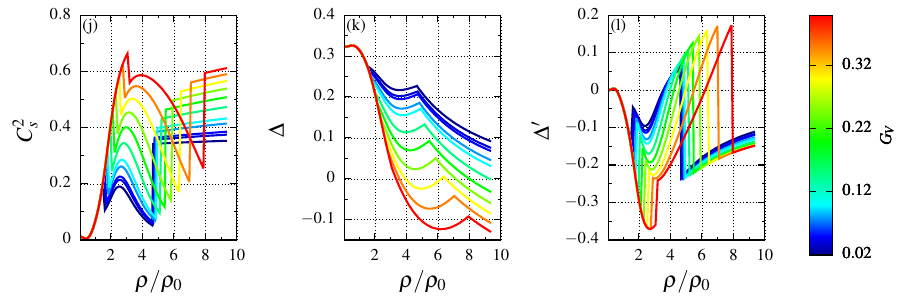}
    \caption{ Speed of sound and its decomposition ($\Delta$ and $\Delta^{\prime}$) versus density in units of the saturation for the temperature , hadronic ($m^*/m$) and  quark parameters($G_V$,$B_0^{1/4}$).}
    \label{fig:cs2_decom}
\end{figure*} 

The effects of the bag constant and the strength of the quark repulsive interaction are
shown in Figures~\ref{fig:part_frac}(e) and \ref{fig:part_frac}(f), respectively.
We find that lower values of the bag constant $B_0^{1/4}$ lead to a more rapid disappearance
of neutrons and a faster growth of the strangeness fraction in the mixed phase.
Similarly, an increase in the repulsive quark interaction strength $G_V$ enhances the
production of strange quarks, resulting in a more rapid rise of the strangeness
fraction within the mixed phase.
In contrast, both $B_0^{1/4}$ and $G_V$ have a relatively weak influence in the pure quark
phase, where the system is already dominated by quark degrees of freedom.
In summary, in the mixed phase, the neutron fraction increases with $m^*/m$, while strangeness is suppressed by Gibbs equilibrium. $K_{\text{sat}}$ has little effect, and symmetry energy mainly impacts the early mixed phase. Lower $B_0^{1/4}$ and higher $G_V$ enhance strangeness, with minimal influence in the pure quark phase.

\begin{figure}[htbp] 
    \centering
   \includegraphics[width=0.15\textwidth]{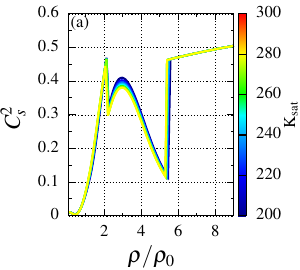}
   \includegraphics[width=0.15\textwidth]{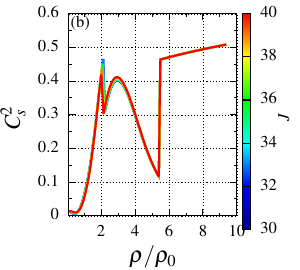}
   \includegraphics[width=0.15\textwidth]{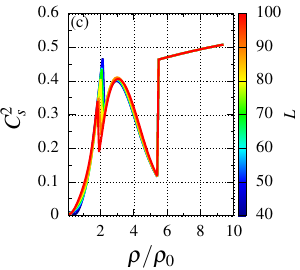}
    \caption{ Variation of the speed of sound as a function of the density normalized to the saturation density for different values of the compressibility($K_{\rm sat}$), symmetry energy($J$) and  its slope ($L$).}
    \label{fig:cs2_nmp}
\end{figure} 
\textbf{\textit{Behavior of entropy per baryon :}}
In Fig.~\ref{fig:entropy}, we have plotted the entropy per baryon ($S/\rho$). To explain the figure, let us first 
consider a fixed temperature, say $T=10$ MeV.  If we change the vector 
interaction ten times from $G_V=0.02$ to $G_V=0.2$, we find that in the quark phase 
the entropy per baryon $s/\rho$ remains unaffected by the vector interaction. 
The reason is as follows: the entropy density depends on the distribution 
function, which in turn depends on the effective chemical potential, the 
quark mass, and the temperature of each species.  The vector 
interaction modifies the quark chemical potential as 
$\mu_i = \mu_i^* + g_V V_0$. In the case of $\beta$-equilibrium, the term $g_V V_0$ is the same for each 
flavor, so it cancels out in the evaluation of the equation of state. 
Effectively, one finds $\mu_d^* = \mu_s^*$, and therefore the entropy 
per baryon is unaffected by the vector interaction.
This can also be understood from the Euler relation. Both the energy 
density and the pressure acquire the same additional contribution, 
$\frac{1}{2} m_V^2 V_0^2.$
When combined, they give $m_V^2 V_0^2$. However, a compensating negative 
contribution arises from the chemical potential term and they cancel exactly. 
Thus, the 
entropy per baryon remains independent of the vector interaction.

In the mixed phase, the quark volume fraction $\chi$ is determined by the 
Gibbs conditions, which require the equality of pressure and chemical potential 
between the hadronic and quark sectors. 
 The quark pressure and chemical potential contains an 
additional contribution from the repulsive vector interaction, which increases 
the stiffness of the quark equation of state. As a result, the pressure balance 
between hadronic and quark matter shifts with the vector coupling $G_V$, and 
therefore the equilibrium value of $\chi$ at a given $(\mu_B,T)$ depends 
explicitly on $G_V$.
The key result is that the entropy per baryon is independent of $G_V$ in the pure quark phase, but in the mixed phase it is affected indirectly through the $G_V$-dependent quark volume fraction $\chi$.

\begin{figure*}[htbp] 
    \centering
    \includegraphics[width=0.99\textwidth]{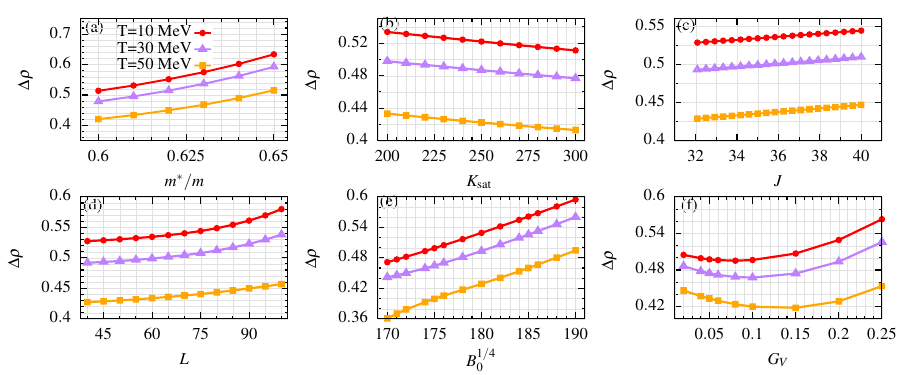}
    \caption{ Sensitivity of the mixed-phase width (in units of $\rm fm^{-3}$) to hadronic and quark sector parameters with three different temperatures (T=10, 30 and 50 MeV). }
    \label{fig:delta_rho}
\end{figure*} 
\textbf{\textit{Speed of sound and its decomposition :}}
The decomposition of the speed of sound in terms of the trace anomaly and its logarithmic
derivative was previously studied in Ref.~\cite{Fujimoto:2022ohj}.
The normalized trace anomaly is defined as
$\Delta = (\varepsilon - 3P)/(3\varepsilon)$. The derivative term is $\Delta^{\prime}=1/3 -\Delta -C_s^2$.
In the present work, we extend this analysis by investigating the behavior of the speed
of sound and its thermodynamic decomposition at finite temperature.

We first examine the temperature dependence of the speed of sound.
In Fig.~\ref{fig:cs2_decom}(a), we observe that with increase in temperature,
the speed of sound in the hadronic phase increases due to
the enhancement of thermal pressure, which steepens the pressure–energy density
relation.
However, in the mixed phase, the speed of sound is reduced as a consequence of the Gibbs
phase equilibrium conditions.
As the temperature increases, the mixed phase adjusts by increasing the quark volume
fraction in order to maintain Gibbs phase equilibrium, which limits the growth of
pressure with energy density. This leads to 
softening of the equation of state and a corresponding reduction of the speed of sound in the mixed phase.
This behavior is also reflected in the evolution of the trace anomaly in Fig.~\ref{fig:cs2_decom}(b).
In the mixed phase, higher temperatures correspond to larger values of the trace anomaly,
indicating stronger deviations from conformal behavior due to the coexistence of
hadronic and quark degrees of freedom under Gibbs phase equilibrium.
At sufficiently high densities, where the system enters the pure quark phase, the trace
anomaly decreases and can become negative, reflecting the increasing dominance of
quark matter and the stiffening of the quark equation of state.
Additional insight is provided by the derivative contribution of the speed of sound (Fig.~\ref{fig:cs2_decom}(c)),
quantified by $\Delta^{\prime}$.
The quantity $\Delta^{\prime}$ can assume both positive and negative values, depending on
the thermodynamic regime. For the hadronic phase, $\Delta^{\prime}$ is always negative (for the details of the thermodynamic decomposition refer to \cite{Pal:2025jln}.)
In the pure hadronic phase, lower temperatures are associated with smaller values of
$\Delta^{\prime}$.
In contrast, within the mixed phase, $\Delta^{\prime}$ exhibits a sign change;
increasing temperature leads to an enhancement of $\Delta^{\prime}$.

We have systematically studied the behavior of the speed of sound for different nuclear and quark matter parameters, as shown in Figs.~\ref{fig:cs2_decom} and ~\ref{fig:cs2_nmp}.
We find that $C_{s}^{2}$ is strongly affected by $m^*/m$, $B_0$, and $G_V$, whereas the influence of $K_{\rm sat}$, $J$, and $L$ is comparatively weak (see Fig.~\ref{fig:cs2_nmp}). Consequently, we do not include the decomposition of the speed of sound for variations of $K_{\rm sat}$, $J$, and $L$ in Fig.~\ref{fig:cs2_nmp}.

We  have  studied  thoroughly the width of the mixed phase, $\Delta \rho$ in Fig.~\ref{fig:delta_rho}, by exploring the
effects of both hadronic and quark matter parameters, as well as thermal effects.
We find that $\Delta \rho$ is influenced by the effective mass ratio $m^*/m$, exhibiting an approximately quadratic dependence on this parameter. $\Delta \rho$  decreases with increasing temperature for a fixed value of the parameter for all the six parameters studied in this figure. 
The effect of nuclear incompressibility in Fig.~\ref{fig:delta_rho}(b) is relatively weak.
Specifically, $\Delta \rho$ decreases with increasing $K_{\text{sat}}$, and this
decrease is approximately linear.

We further investigate the impact of the symmetry energy and its slope on the mixed
phase width.
The symmetry energy coefficient $J$ has a noticeable effect on $\Delta \rho$, with an
almost linear dependence.
The slope of the symmetry energy, $L$, also affects the mixed phase width; although the
variation is not strictly linear, its influence becomes more pronounced at larger
values of $L$.
Overall, an increase in both $J$ and $L$ leads to a widening of the mixed phase region.

To illustrate the role of quark matter parameters and temperature more clearly, we
present in Fig.~\ref{fig:delta_rho}(e) and (f) the variation of the mixed-phase width,
$\Delta \rho$, as a function of the vector coupling strength $G_V$, the bag constant
$B_0^{1/4}$, and the temperature $T$.
We find that the mixed-phase width depends approximately linearly on the bag constant,
increasing with increase in $B_0^{1/4}$ which  reflects the delayed onset of quark matter at higher values of
bag constants.
In contrast, the dependence of $\Delta \rho$ on the vector coupling strength $G_V$ is
non-monotonic.
For lower values of $G_V$, the mixed-phase width decreases, whereas for larger values of
$G_V$, $\Delta \rho$ increases, resulting in a parabolic-type dependence.
Our analysis indicates that, within a certain range of the vector interaction strength,
the repulsive vector interaction plays a crucial role in enhancing the formation and
stability of the hadron–quark mixed phase.

We now summarize the key result: temperature softens the equation of state in the mixed phase, reducing $C_s^2$ while enhancing the trace anomaly, whereas $m^*/m$, $B_0$, and $G_V$ strongly influence the speed of sound. The mixed-phase width $\Delta \rho$ decreases with temperature, increases with $J$ and $L$, grows linearly with $B_0^{1/4}$, and shows a non-monotonic dependence on $G_V$.

\subsection{Structural properties and astrophysical implication}
Having established the thermodynamic properties of the quark–hadron mixed phase and
their dependence on hadronic and quark matter parameters, we now turn to the
astrophysical implications of our equation of state.
In particular, we investigate how the temperature-dependent hybrid EOS affects the
global properties of neutron stars (considered T=10 MeV). In all figures illustrating the structural properties, the solid lines correspond to a strong repulsive vector coupling, $G_V=0.2~\text{fm}^2$ and dashed lines represent a weaker coupling,
$G_V=0.02~\text{fm}^2$.

\subsubsection{Effect of symmetric nuclear matter parameters}

\begin{figure*}[htbp] 
    \centering
    \includegraphics[width=0.45\textwidth]{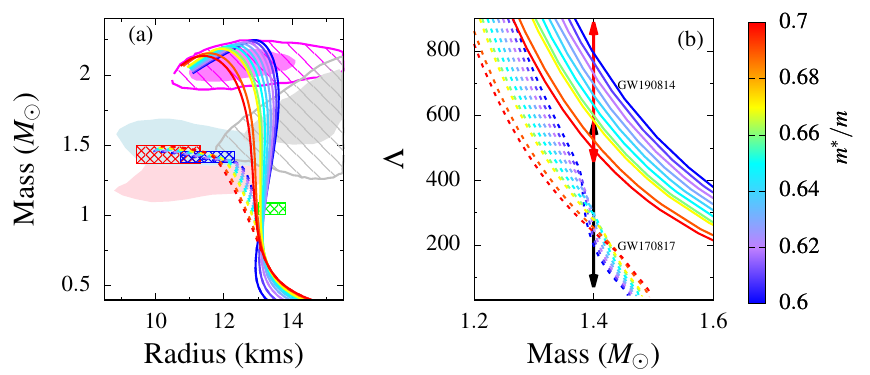}
    \includegraphics[width=0.45\textwidth]{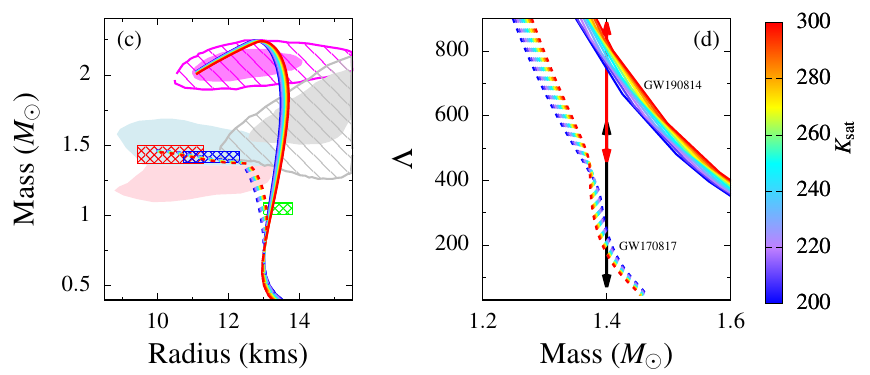}
    \caption{ Mass–radius relations and the corresponding dimensionless tidal deformability for hybrid star configurations (upper panels). Results are shown for different values of $m^*/m$ (left panels) and $K_{\rm sat}$ (right panels).
   Solid lines represent $G_V = 0.2~\mathrm{fm}^2$, while dashed lines represent $G_V = 0.02~\mathrm{fm}^2$.
 }
    \label{fig:mr_mstar_Ksat}
\end{figure*}

\begin{table*}[!ht] 
\centering
\caption{Neutron star properties for different $m^*/m$, $K_{\rm sat}$, and $G_V$. 
$M$ and $R$ denote mass (in $M_\odot$) and radius (in km), $\Lambda$ is the dimensionless tidal deformability. 
$\chi=0$ corresponds to non-rotating stars, $\chi=1$ to maximally rotating stars.}
\label{tab:mstar_ksat} 
\setlength{\tabcolsep}{3pt} 
\begin{tabular}{ccccccccccccc}
\hline
\hline
$m^*/m$ & $G_V$ & $M(\chi=0)$ & $M(\chi=1)$ & $\Delta M$ & $R(\chi=0)$ & $R(\chi=1)$ & $\Delta R$ & $\Lambda(\chi=0)$ & $\Lambda(\chi=1)$ & $\Delta \Lambda$ & $M_{\max}$ & $R_{M_{\max}}$ \\
\hline
0.600 & 0.20 & 1.599 & 2.246 & 0.646 & 13.534 & 12.810 & 0.724 & $3.808\times10^{2}$ & $2.125\times10^{1}$ & $3.595\times10^{2}$ & 2.249 & 12.901 \\
0.653 & 0.20 & 1.541 & 2.198 & 0.657 & 13.195 & 11.985 & 1.210 & $3.618\times10^{2}$ & $1.289\times10^{1}$ & $3.489\times10^{2}$ & 2.206 & 12.164 \\
0.700 & 0.20 & 1.511 & 2.108 & 0.597 & 12.777 & 10.986 & 1.791 & $3.091\times10^{2}$ & 7.605 & $3.015\times10^{2}$ & 2.136 & 11.398 \\
\hline
0.600 & 0.02 & 0.700 & 1.378 & 0.679 & 12.935 & 12.350 & 0.585 & $2.309\times10^{4}$ & $4.193\times10^{2}$ & $2.267\times10^{4}$ & 1.457 & 9.935 \\
0.653 & 0.02 & 0.673 & 1.406 & 0.734 & 13.229 & 11.992 & 1.237 & $2.645\times10^{4}$ & $2.746\times10^{2}$ & $2.618\times10^{4}$ & 1.475 & 10.080 \\
0.700 & 0.02 & 0.653 & 1.445 & 0.792 & 13.271 & 11.517 & 1.754 & $2.879\times10^{4}$ & $1.678\times10^{2}$ & $2.862\times10^{4}$ & 1.494 & 10.210 \\
\hline
$K_{\rm sat}$ & $G_V$ & $M(\chi=0)$ & $M(\chi=1)$ & $\Delta M$ & $R(\chi=0)$ & $R(\chi=1)$ & $\Delta R$ & $\Lambda(\chi=0)$ & $\Lambda(\chi=1)$ & $\Delta \Lambda$ & $M_{\max}$ & $R_{M_{\max}}$ \\
\hline
200 & 0.20 & 1.565 & 2.242 & 0.677 & 13.423 & 12.646 & 0.777 & $3.964\times10^{2}$ & $1.889\times10^{1}$ & $3.775\times10^{2}$ & 2.246 & 12.737 \\
300 & 0.20 & 1.645 & 2.237 & 0.592 & 13.657 & 12.851 & 0.806 & $3.398\times10^{2}$ & $2.209\times10^{1}$ & $3.177\times10^{2}$ & 2.241 & 12.950 \\
\hline
200 & 0.02& 0.6786 & 1.3876 & 0.7090 & 13.0071 & 12.2411 & 0.7660 &
$2.583\times10^{4}$ & $3.675\times10^{2}$ & $2.546\times10^{4}$ &
1.4630 & 9.9571 \\
300 & 0.02 &0.7324 & 1.3713 & 0.6389 & 12.9961 & 12.4741 & 0.5220 &
$1.927\times10^{4}$ & $4.619\times10^{2}$ & $1.881\times10^{4}$ &
1.4522 & 9.9361 \\
\hline
\hline
\end{tabular}
\end{table*}

Fig.~\ref{fig:mr_mstar_Ksat} shows the impact of the symmetric nuclear matter parameters,
$m^*/m$ and $K_{\text{sat}}$, on hybrid star properties for weak ($G_V=0.02$) and strong
($G_V=0.2$) vector interactions. Panels (a,c) display the mass--radius relations, (b,d)
the tidal deformabilities.
We impose observational constraints on the stellar radius from the massive pulsar PSR~J0740+6620 ($10.08$–$12.49$ km, shown in magenta), the $1.4\,M_{\odot}$ pulsar PSR~J0030+0451 ($11.96$–$14.26$ km, shown in grey), and the solar-mass pulsar PSR~J1231–1411 ($13.0$–$13.8$ km, shown in green).


\textit{Mass--radius relation ($M$--$R$):}
The mass--radius relation shows a strong dependence on $m^*/m$, while the effect of
$K_{\text{sat}}$ is relatively weak. For $G_V = 0.2$, decreasing $m^*/m$ increases the
maximum mass and yields configurations compatible with PSR~J0740+6620, PSR~J0030+0451 and PSR~J1231--1411, whereas the canonical mass and low-radius constraints
from PSR~J0437--4715 and PSR~J0614--3329 are satisfied only for $G_V = 0.02$. Variations in
$K_{\text{sat}}$ mainly lead to a reordering of the mass--radius curves after the onset
of the quark--hadron phase transition, with only small changes in the maximum mass.

\textit{Tidal deformability ($\Lambda$--$M$):}
The dimensionless tidal deformability depends sensitively on $m^*/m$ for the hadronic
branch, whereas the quark branch shows only a weak dependence. For $G_V = 0.2$, lower
values of $m^*/m$ satisfy the GW190814 constraint\cite{LIGOScientific:2020zkf}($\Lambda_{1.4M_{\odot}}=616_{-158}^{+273}$) , while intermediate values
($m^*/m \simeq 0.68$--$0.70$) are consistent with both GW170817 \cite{LIGOScientific:2018cki} ($\Lambda_{1.4M_{\odot}}=190_{-120}^{+390}$) and GW190814. For
$G_V = 0.02$, the GW170817 constraints are satisfied for all considered values of
$m^*/m$. A similar reordering is observed in the $\Lambda$--$M$ relation when varying
$K_{\text{sat}}$.


\color{black}

\begin{figure*}[htbp] 
    \centering
    \includegraphics[width=0.45\textwidth]{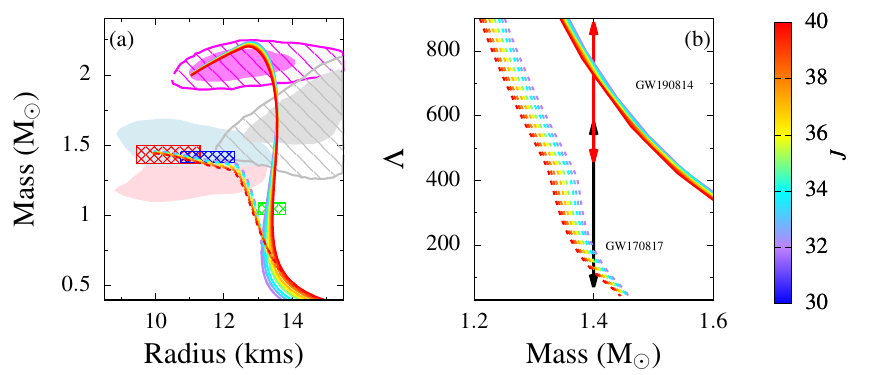}
    \includegraphics[width=0.45\textwidth]{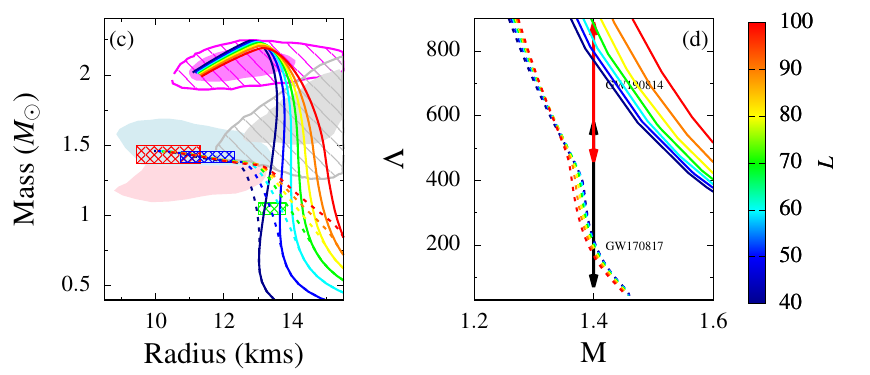}
    \caption{ Same as Fig.~\ref{fig:mr_mstar_Ksat}, but for different values of the symmetry energy $J$ (left panels) and its slope parameter $L$(right panels).
 }
    \label{fig:mrlam_JL}
\end{figure*}

\begin{table*}[!ht]
\centering
\caption{Same as Table.~\ref{tab:mstar_ksat}, but for the different $J$, $L$ and $G_V$.}
\label{tab:J_L} 
\setlength{\tabcolsep}{2.0 pt}
\begin{tabular}{ccccccccccccc}
\hline
\hline 
 & $G_V$ & $M(\chi=0)$ & $M(\chi=1)$ & $\Delta M$ &
$R(\chi=0)$ & $R(\chi=1)$ & $\Delta R$ &
$\Lambda(\chi=0)$ & $\Lambda(\chi=1)$ & $\Delta\Lambda$ &
$M_{\max}$ & $R_{M_{\max}}$ \\
\hline
J=32 & 0.02 &
0.6634 & 1.3728 & 0.7094 &
13.1161 & 12.3051 & 0.8110 &
$2.939\times10^{4}$ & $4.016\times10^{2}$ & $2.899\times10^{4}$ &
1.4571 & 9.9471 \\

J=40 & 0.02 &
0.4829 & 1.3240 & 0.8411 &
14.2161 & 12.1951 & 2.0210 &
$1.422\times10^{5}$ & $4.289\times10^{2}$ & $1.418\times10^{5}$ &
1.4439 & 9.8841 \\
J=32 & 0.20 &
1.5621 & 2.2344 & 0.6722 &
13.5051 & 12.7001 & 0.8050 &
$4.150\times10^{2}$ & $1.994\times10^{1}$ & $3.951\times10^{2}$ &
2.2376 & 12.7931 \\
J=40 & 0.20 &
1.3810 & 2.2009 & 0.8199 &
13.5161 & 12.6401 & 0.8760 &
$7.769\times10^{2}$ & $2.048\times10^{1}$ & $7.564\times10^{2}$ &
2.2029 & 12.7151 \\
\hline 
L=40 & 0.02 &
0.6959 & 1.3817 & 0.6858 &
13.0241 & 12.3191 & 0.7050 &
$2.348\times10^{4}$ & $3.950\times10^{2}$ & $2.309\times10^{4}$ &
1.4596 & 9.9551 \\

L=95 & 0.02 &
0.8001 & 1.3619 & 0.5617 &
15.4461 & 13.0101 & 2.4360 &
$2.396\times10^{4}$ & $4.507\times10^{2}$ & $2.351\times10^{4}$ &
1.4529 & 10.2681 \\
L=40 & 0.20 &
1.5894 & 2.2400 & 0.6506 &
13.5041 & 12.7081 & 0.7960 &
$3.782\times10^{2}$ & $1.978\times10^{1}$ & $3.584\times10^{2}$ &
2.2436 & 12.8011 \\

L=95 & 0.20 &
1.4190 & 2.1983 & 0.7793 &
14.7921 & 13.0331 & 1.7590 &
$1.007\times10^{3}$ & $2.196\times10^{1}$ & $9.853\times10^{2}$ &
2.2006 & 13.1191 \\
\hline
\hline 
\end{tabular}
\end{table*}

\subsubsection{Effect of symmetry energy and its slope parameters}

In addition to constraints from astrophysical observations, the nuclear symmetry energy parameters $J$ and $L$ are also constrained by experiments and theoretical studies. 
Constraints on the $J$--$L$ correlation have been obtained from a variety of experimental measurements and theoretical approaches~\cite{Drischler:2020hwi}. 
In particular, the PREX collaboration ~\cite{Reed:2021nqk}. exploited the strong correlation between the neutron-skin thickness of ${}^{208}\mathrm{Pb}$ and the slope of the nuclear symmetry energy within a class of relativistic energy density functionals, providing estimates for $J$ and $L$ that tend to favor relatively large values of the symmetry energy slope. 
Motivated by these experimental and theoretical considerations, in this work we vary the symmetry energy parameters over the ranges $J = 30$--$40$ MeV and $L = 40$--$100$ MeV.
In this subsection, we investigate the impact of the symmetry energy parameter $J$ and $L$ on the global properties of compact stars and on the characteristics of the hadron--quark phase transition. We first present the mass--radius and mass--tidal deformability relations, followed by a detailed analysis of the transition-induced changes in mass, radius, and internal composition, as discussed in the previous subsection.

Figures~\ref{fig:mrlam_JL}(a-d)--upper panel display the mass--radius and mass--tidal deformability relations, first two columns for $J$ and the 3rd and 4th columns
for $L$. The details of the structural properties for $J$ and $L$ is given also in the table.~\ref{tab:J_L}. 

\textit{Mass--radius relation ($M$--$R$):}  
Figures~\ref{fig:mrlam_JL}(a) show the mass--radius relations for varying $J$, while Figures~\ref{fig:mrlam_JL}(c) correspond to variations in $L$. For both strong ($G_V = 0.2$) and weak ($G_V = 0.02$) quark-matter repulsion, increasing $J$ or $L$ generally enlarges the hadronic branch radius, whereas the quark-star branch remains comparatively insensitive.  
From an observational perspective, both symmetry energy parameters strongly influence the stellar radius in the strong repulsion case ($G_V = 0.2$). For different $J$ values, the predicted mass--radius relations are fully consistent with the NICER constraint from PSR~J1231$-$1411, as indicated by the overlap with the green region in the $M$--$R$ plane. Since the stellar radius is more sensitive to $L$, lower values of $L$ are favored by the PSR~J1231$-$1411 observations. Additionally, for $G_V = 0.2$, the NICER constraint from PSR~J0030$+$0451 is also fully satisfied.  
The behavior of the maximum mass and the corresponding radius is summarized in Table~\ref{tab:J_L} for variations of $J$ and $L$ under both weak and strong quark-matter repulsion. In both interaction scenarios, the qualitative impact of $J$ and $L$ on the maximum mass is similar: increasing either parameter leads to a mild reduction in $M_{\max}$, while the corresponding radius changes only moderately.

\textit{Tidal deformability ($\Lambda$--$M$):}  
The corresponding mass--tidal deformability ($M$--$\Lambda$) relations are shown in Fig.~\ref{fig:mrlam_JL}(b) for variations in $J$ and in Fig.~\ref{fig:mrlam_JL}(d) for variations in $L$.  
For the strong quark-matter repulsion case ($G_V = 0.2$), the tidal deformability of the purely hadronic configuration, $\Lambda(\chi = 0)$, increases noticeably with increasing $J$, whereas the deformability of the fully converted quark configuration, $\Lambda(\chi = 1)$, shows only a mild variation. A similar trend is observed for the weakly repulsive interaction ($G_V = 0.02$), although the increase in $\Lambda(\chi = 0)$ becomes much more pronounced due to the larger hadronic radius. In contrast, the quark-star branch remains nearly insensitive to changes in $J$. 
The tidal deformability of the purely hadronic configuration, $\Lambda(\chi = 0)$, increases rapidly with $L$, reflecting the strong correlation between the symmetry energy slope and the stellar radius. 
In contrast, the tidal deformability of the fully converted quark configuration, $\Lambda(\chi = 1)$, exhibits only a weak dependence on $L$.

\begin{figure*}[htbp] 
    \centering
    \includegraphics[width=0.45\textwidth]{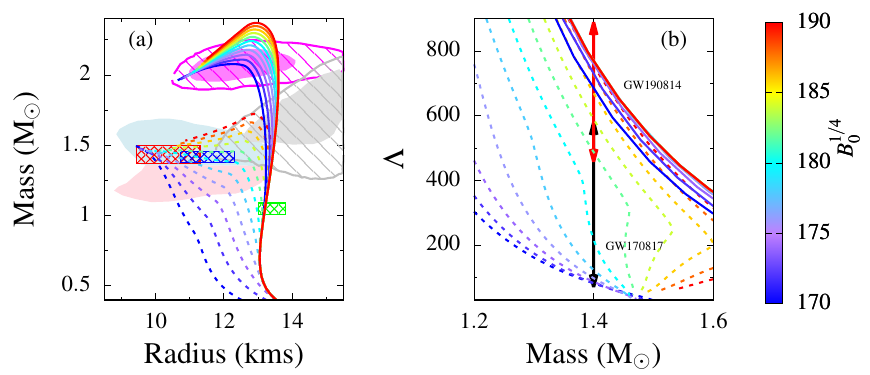}
    \includegraphics[width=0.45\textwidth]{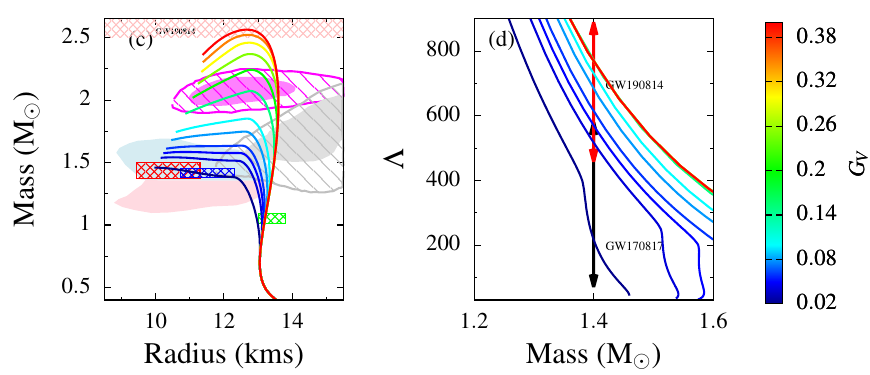}
        \includegraphics[width=0.45\textwidth]{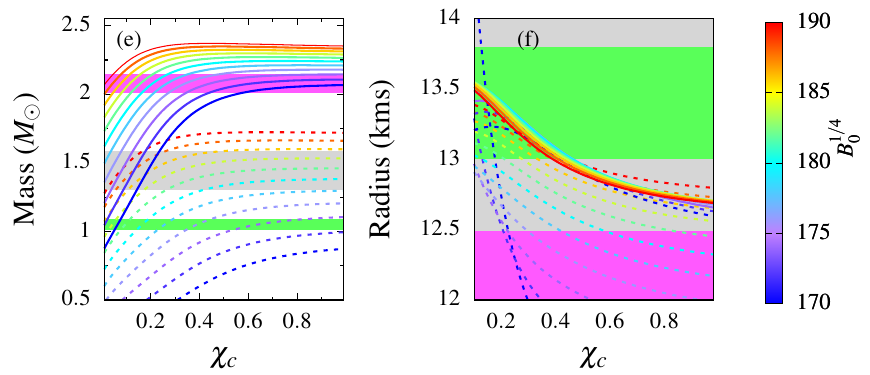}
   \includegraphics[width=0.45\textwidth]{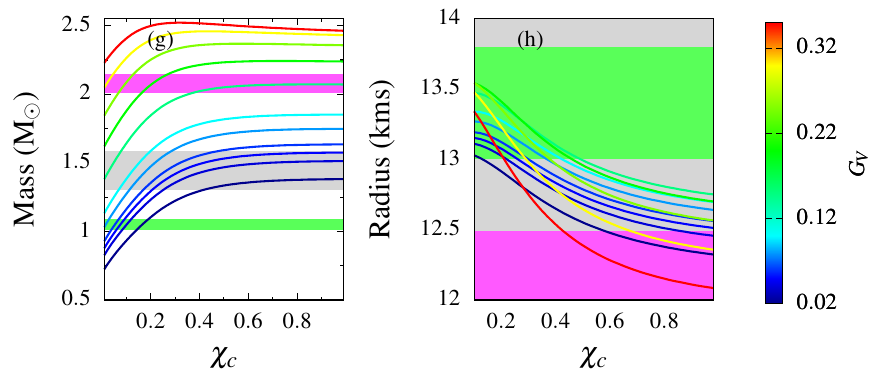}
        \includegraphics[width=0.45\textwidth]{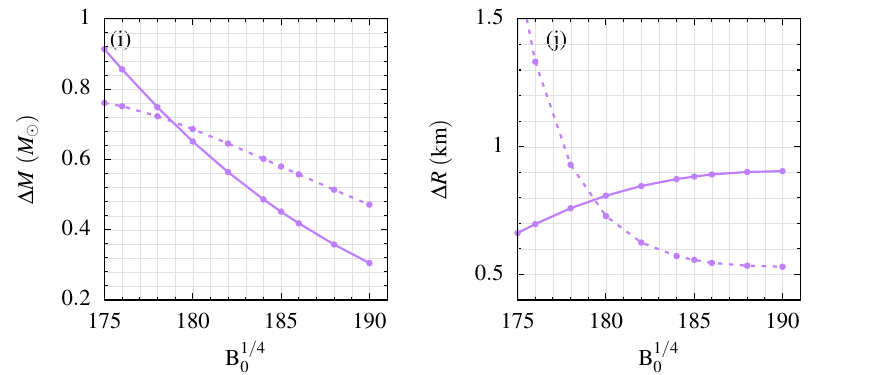}
     \includegraphics[width=0.45\textwidth]{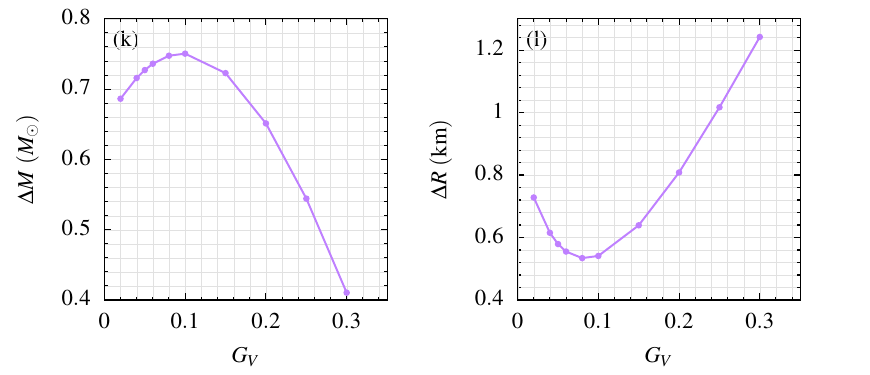}
    \caption{ Mass–radius relations and the corresponding dimensionless tidal deformability for hybrid star configurations (upper
panels). Variation of the stellar mass with the quark volume fraction and variation of the stellar radius with the quark volume
fraction in the mixed phase (middle panels). Dependence of the mass and radius differences across the hadron–quark phase
transition (lower panels) for different values of the  $B_0^{1/4}$ (left panels) and  $G_V$(right panels).}
    \label{fig:mrl_B14GV}
\end{figure*} 

\begin{figure}[htbp] 
    \centering
       \includegraphics[width=0.45\textwidth]{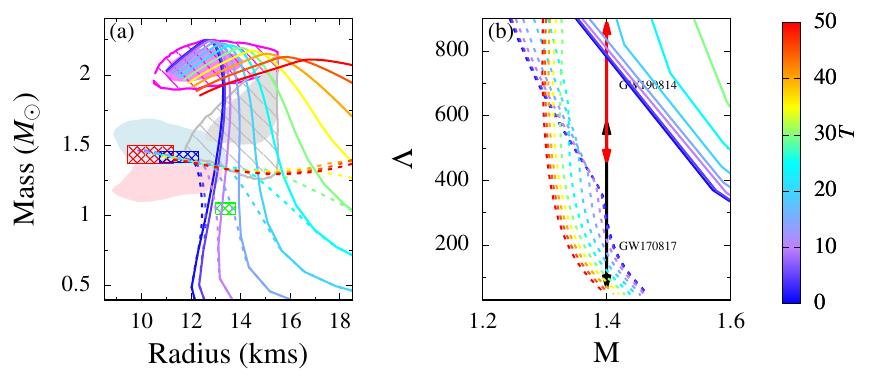}
    \includegraphics[width=0.45\textwidth]{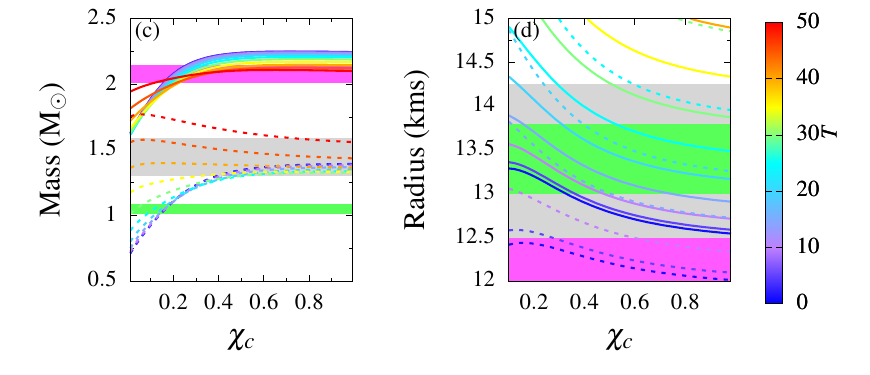}
\includegraphics[width=0.45\textwidth]{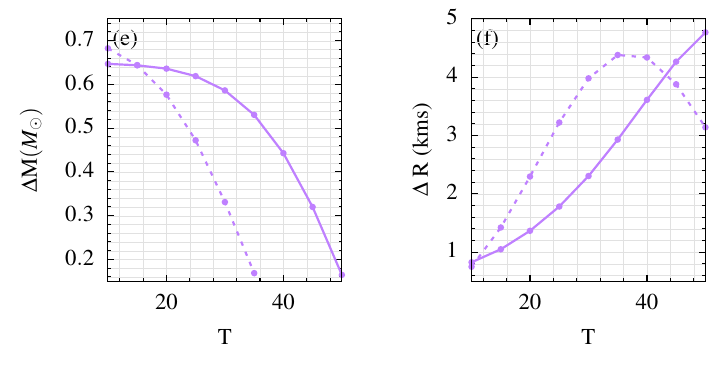}
    \caption{Same as Fig.~\ref{fig:mr_mstar_Ksat}, but showing finite-temperature effects on the structural properties of hybrid stars for two different vector coupling strengths.}
    \label{fig:mrl_T}
\end{figure}



\begin{table*}[!ht]
\centering
\caption{Same as Table.~\ref{tab:mstar_ksat}, but for the different $B_0^{1/4}$ and $G_V$.}
\label{tab:transition_J} 
\setlength{\tabcolsep}{2.0pt}
\begin{tabular}{ccccccccccccc}
\hline
$B_0^{1/4}$ & $G_V$ & $M(\chi=0)$ & $M(\chi=1)$ & $\Delta M$ &
$R(\chi=0)$ & $R(\chi=1)$ & $\Delta R$ &
$\Lambda(\chi=0)$ & $\Lambda(\chi=1)$ & $\Delta\Lambda$ &
$M_{\max}$ & $R_{M_{\max}}$ \\
\hline
172 & 0.20 &
0.8428 & 2.0673 & 1.2245 &
13.0851 & 12.5901 & 0.4950 &
$9.706\times10^{3}$ & $3.415\times10^{1}$ & $9.672\times10^{3}$ &
2.0778 & 12.2451 \\

175 & 0.20 &
1.2486 & 2.1622 & 0.9136 &
13.3251 & 12.6631 & 0.6620 &
$1.405\times10^{3}$ & $2.527\times10^{1}$ & $1.380\times10^{3}$ &
2.1623 & 12.6751 \\

190 & 0.20 &
2.0463 & 2.3517 & 0.3054 &
13.5911 & 12.6871 & 0.9040 &
$7.672\times10^{1}$ & $1.321\times10^{1}$ & $6.350\times10^{1}$ &
2.3711 & 12.9371 \\
\hline
172 & 0.02 &
0.1659 & 0.8740 & 0.7081 &
21.6611 & 11.1441 & 10.5170 &
$2.719\times10^{7}$ & $2.805\times10^{3}$ & $2.718\times10^{7}$ &
1.5012 & 9.2911 \\

175 & 0.02 &
0.3955 & 1.1563 & 0.7608 &
13.5581 & 11.8751 & 1.6830 &
$3.060\times10^{5}$ & $8.680\times10^{2}$ & $3.051\times10^{5}$ &
1.4664 & 9.5831 \\

190 & 0.02 &
1.2478 & 1.7191 & 0.4713 &
13.3251 & 12.7951 & 0.5300 &
$1.410\times10^{3}$ & $1.365\times10^{2}$ & $1.274\times10^{3}$ &
1.7231 & 12.8731 \\
\hline\hline
180 & 0.02 &
0.6951 & 1.3815 & 0.6864 &
13.0471 & 12.3191 & 0.7280 &
$2.377\times10^{4}$ & $3.950\times10^{2}$ & $2.338\times10^{4}$ &
1.4596 & 9.9591 \\

180 & 0.05 &
0.8478 & 1.5749 & 0.7271 &
13.0871 & 12.5081 & 0.5790 &
$9.438\times10^{3}$ & $1.994\times10^{2}$ & $9.239\times10^{3}$ &
1.5847 & 10.8701 \\

180 & 0.10 &
1.1010 & 1.8515 & 0.7505 &
13.2331 & 12.6921 & 0.5410 &
$2.663\times10^{3}$ & $7.814\times10^{1}$ & $2.584\times10^{3}$ &
1.8515 & 12.6921 \\

180 & 0.15 &
1.3489 & 2.0717 & 0.7228 &
13.3851 & 12.7461 & 0.6390 &
$9.367\times10^{2}$ & $3.673\times10^{1}$ & $8.999\times10^{2}$ &
2.0723 & 12.7801 \\

180 & 0.20 &
1.5876 & 2.2388 & 0.6512 &
13.5041 & 12.6961 & 0.8080 &
$3.810\times10^{2}$ & $1.970\times10^{1}$ & $3.613\times10^{2}$ &
2.2424 & 12.7901 \\

180 & 0.25 &
1.8129 & 2.3571 & 0.5442 &
13.5771 & 12.5601 & 1.0170 &
$1.719\times10^{2}$ & $1.169\times10^{1}$ & $1.603\times10^{2}$ &
2.3686 & 12.7691 \\

180 & 0.30 &
2.0203 & 2.4307 & 0.4104 &
13.5941 & 12.3521 & 1.2420 &
$8.397\times10^{1}$ & $7.531$ & $7.644\times10^{1}$ &
2.4590 & 12.7411 \\
\hline
\end{tabular}
\end{table*}


\subsection{Effect of quark matter parameters}
In this subsection, we investigate the impact of the quark-matter model parameters, namely the bag constant $B_0^{1/4}$ and the vector coupling strength $G_V$, on  the global properties of compact stars, as well as on the characteristics of the hadron--quark phase transition.
From an astrophysical perspective, both parameters play a crucial role in determining the compatibility of hybrid star configurations with observational constraints, including those inferred from NICER and gravitational-wave observations.

\textit{Mass--radius relation ($M$--$R$):}  
Figure~\ref{fig:mrl_B14GV}(a) shows the mass--radius relations for different values of the bag parameter $B_0^{1/4}$, considering both weak ($G_V = 0.02$) and strong ($G_V = 0.2$) vector couplings.
For the weakly repulsive case, represented by the dashed curves, lower values of $B_0^{1/4}$ (up to $\sim 175$ MeV) fail to satisfy the NICER constraint from PSR~J1231$-$1411 (green rectangle), while remaining  are compatible with the mass--radius regions inferred for PSR~J0614$-$3329 and PSR~J0437$-$4715.
As $B_0^{1/4}$ increases, the predicted radii decrease, and the mass--radius relations shift into the region allowed by PSR~J1231$-$1411 and PSR~J0030$+$0451.
However, in this regime, the constraints from PSR~J0614$-$3329 and PSR~J0437$-$4715 are no longer satisfied, indicating a tension between different observational bounds in the weak vector interaction scenario.
In contrast, for strong vector coupling ($G_V = 0.2$), shown by the solid curves, the mass--radius relations are systematically shifted toward larger radii for a given mass.
For the selected range of $B_0^{1/4}$, the strong-repulsion scenario allows simultaneous agreement with the NICER constraints from PSR~J1231$-$1411, PSR~J0030$+$0451, and PSR~J0740$+$6620.
Nevertheless, owing to the enhanced repulsive interaction, it remains challenging to reproduce all canonical pulsar constraints simultaneously across the entire parameter space.

Figure~\ref{fig:mrl_B14GV}(c) further illustrates the role of the vector coupling strength. The vector coupling plays a crucial role in reconciling theoretical models with astrophysical observations over a wide mass range, from low-mass to high-mass pulsars. In particular, compact stars with relatively small radii at $1.4,M_\odot$ can be interpreted as hybrid stars with weak vector repulsion, which naturally produce softer equations of state and more compact configurations.

\textit{Tidal deformability ($\Lambda$--$M$):}  
The corresponding mass--tidal deformability relations are shown in Fig.~\ref{fig:mrl_B14GV}(b) and (d).
For the weakly repulsive case ($G_V = 0.02$), the constraints from GW170817 are satisfied up to intermediate values of the bag constant, $B_0^{1/4} \lesssim 175$ MeV but for the higher values of  $B_0^{1/4}$,  GW190814 constraint is satisfied.
In contrast, for stronger vector coupling ($G_V = 0.2$), the predicted tidal deformabilities exceed the GW170817 bounds, and only the looser constraints inferred from GW190814 remain satisfied.
From a broader perspective, smaller values of $G_V$ are favored by tidal deformability observations, as they lead to more compact stars and reduced $\Lambda$, whereas larger $G_V$ values allow the support of very massive stars at the cost of increased tidal deformabilities.

\begin{figure}[htbp] 
    \centering
       \includegraphics[width=0.45\textwidth]{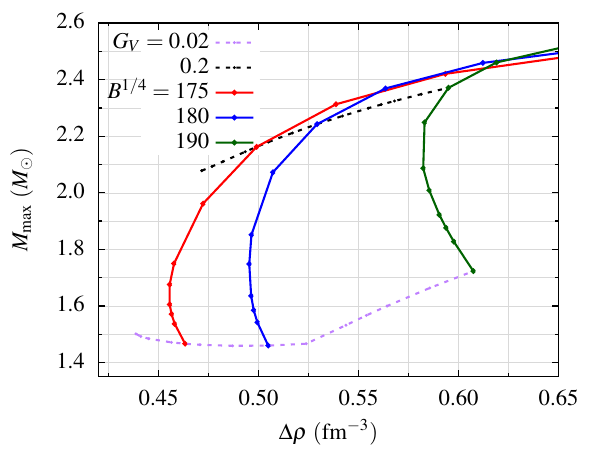}
  \caption{Maximum gravitational mass ($M_{\mathrm{max}}$) as a function of the density extension of the mixed phase ($\Delta \rho$) for variations of  the quark parameters.}
    \label{fig:M_max_delta_rho}
\end{figure} 
\textit{Mixed-phase mass and radius ($M$--$\chi_c$, $R$--$\chi_c$):}
For strong vector interaction ($G_V = 0.2$), increasing $B_0^{1/4}$ raises both the quark-matter onset mass $M(\chi = 0)$ and the fully quark-dominated mass $M(\chi = 1)$, indicating enhanced stability of the hadronic phase.
The corresponding $R$–$\chi_c$ relation shows a comparatively weaker dependence on $B_0^{1/4}$. For $G_V=0.2$, the hadronic radius at the transition, $R(\chi=0)$, increases mildly with $B_0^{1/4}$, while the quark-star radius, $R(\chi=1)$, remains nearly constant around $\sim 12.6$–$12.7$ km. This implies that changes in the bag constant primarily affect the mass evolution with $\chi_c$, whereas the stellar radius is less sensitive once a sizable quark core is present.

In contrast, for the weakly repulsive case ($G_V=0.02$), the $M$–$\chi_c$ relation exhibits a much stronger sensitivity to $B_0^{1/4}$. At low bag constants, the onset mass $M(\chi=0)$ is very small, indicating an early appearance of quark matter, while increasing $B_0^{1/4}$ shifts the transition to higher masses and significantly increases both $M(\chi=1)$ and the maximum mass. The associated $R$–$\chi_c$ behavior is also more dramatic: for low $B_0^{1/4}$, the hadronic radius at the transition can be extremely large, whereas higher values of $B_0^{1/4}$ lead to more compact and astrophysically realistic radii.

The $M$–$\chi_c$ relation indicates that increasing $G_V$ systematically shifts the onset of the hadron–quark phase transition to higher stellar masses. For weak vector repulsion ($G_V=0.02$), quark matter appears at relatively low masses, with $M(\chi=0)\simeq0.70,M_\odot$, and the growth of the quark core rapidly softens the equation of state, limiting the maximum mass to $M_{\max}\simeq1.46,M_\odot$. As $G_V$ increases, the repulsive interaction delays the formation of quark matter, leading to a monotonic increase in both $M(\chi=0)$ and $M(\chi=1)$. Consequently, the maximum mass rises significantly, exceeding $2,M_\odot$ for $G_V\gtrsim0.15$ and reaching $\sim2.46,M_\odot$ at $G_V=0.30$.
The corresponding $R$–$\chi_c$ relation exhibits a milder but systematic dependence on $G_V$. For small values of the vector coupling, the stellar radius decreases noticeably as the quark fraction increases, reflecting the softening of the equation of state induced by quark matter. With increasing $G_V$, both the hadronic radius at the transition, $R(\chi=0)$, and the quark-core radius, $R(\chi=1)$, increase gradually, indicating a stiffer high-density equation of state.

\textit{Mass and radius differences ($\Delta M$, $\Delta R$):}
The mass difference decreases with increasing $B_0^{1/4}$ for both strong and repulsive vector interactions, while the radius difference $\Delta R$ increases with $B_0^{1/4}$ for $G_V = 0.2$ but decreases for weaker vector coupling.
The mass difference across the transition, $\Delta M$, exhibits a non-monotonic dependence on the vector coupling strength. It increases with $G_V$ at low coupling, reaching a maximum at intermediate values ($G_V\simeq0.10$), and then decreases for stronger repulsive interactions. In contrast, the radius difference, $\Delta R$, initially decreases with increasing $G_V$, attaining a minimum at intermediate coupling, and subsequently increases for larger $G_V$.

\begin{table*}[!ht]
\centering
\caption{Temperature dependence of the hadron--quark phase transition and global stellar properties for fixed model parameters.}
\label{tab:T_dependence}
\setlength{\tabcolsep}{4pt}
\begin{tabular}{ccccccccccccc}
\hline
$G_V$ & $T$& $M(\chi=0) $ & $M(\chi=1) $ & $\Delta M$ & $R(\chi=0) $ & $R (\chi=1)$ & $\Delta R$ & $\Lambda (\chi=0) $ & $\Lambda (\chi=1) $ & $\Delta \Lambda$ & $M_{\max}$ & $R_{M_{\max}}$ \\
\hline
0.20 & 10 & 1.593 & 2.240 & 0.647 & 13.54 & 12.71 & 0.83 & $3.61\times10^{2}$ & $1.91\times10^{1}$ & $3.42\times10^{2}$ & 2.244 & 12.80 \\
0.20 & 30 & 1.604 & 2.190 & 0.586 & 16.17 & 13.87 & 2.31 & $6.15\times10^{2}$ & $2.83\times10^{1}$ & $5.87\times10^{2}$ & 2.193 & 13.98 \\
0.20 & 50 & 1.938 & 2.102 & 0.165 & 21.19 & 16.42 & 4.77 & $1.01\times10^{3}$ & $8.25\times10^{1}$ & $9.24\times10^{2}$ & 2.111 & 16.86 \\
\hline
\hline
0.02 & 10 & 0.700 & 1.383 & 0.683 & 13.08 & 12.33 & 0.75 & $2.23\times10^{4}$ & $3.82\times10^{2}$ & $2.19\times10^{4}$ & 1.460 & 9.97 \\
0.02 & 30 & 0.998 & 1.329 & 0.331 & 18.83 & 14.85 & 3.98 & $1.93\times10^{4}$ & $8.36\times10^{2}$ & $1.85\times10^{4}$ & 1.429 & 10.64 \\
0.02 & 50 & 1.748 & 1.560 & $-0.188$ & 23.94 & 20.80 & 3.14 & $4.76\times10^{3}$ & $2.72\times10^{3}$ & $2.05\times10^{3}$ & 1.772 & 23.50 \\
\hline
\end{tabular}
\end{table*}

\subsection{Finite-temperature effects on the structural properties of hybrid stars}
Figure~\ref{fig:mrl_T} summarizes the impact of finite temperature on the global properties of hybrid stars for weak ($G_V=0.02$) and strong ($G_V=0.2$) vector interactions.
Increasing temperature shifts the $M$–$R$ curves toward larger radii for a given mass, indicating thermal expansion of the stellar configuration. This effect is present for both values of $G_V$, but stars with stronger vector repulsion remain more compact at the same mass.
For $G_V=0.02$, lower temperatures satisfy the GW170817 constraint, while increasing temperature leads to larger tidal deformabilities due to radius inflation. For $G_V=0.2$, the deformability remains smaller, consistent with heavier and more compact stars
The maximum mass is reached at smaller $\chi_c$ as temperature increases, implying that thermal effects promote earlier quark-core formation but reduce the ability of the star to support additional mass.
The stellar radius increases monotonically with temperature across the full range of $\chi_c$, consistent with the behavior seen in the $M$–$R$ plane.
As shown in Table~\ref{tab:T_dependence}, $\Delta M$ decreases with temperature for both vector couplings, reflecting a reduced mass jump across the hadron–quark transition at higher temperatures.
In contrast, $\Delta R$ increases with temperature, reaching $\sim4$–$5$ km at $T=50$ MeV, due to the strong thermal expansion of the hadronic phase relative to the quark phase.
Overall, finite-temperature effects soften the equation of state, leading to larger stellar radii and a systematic reduction of the maximum supported mass, while strong vector repulsion mitigates these effects and preserves compatibility with massive pulsar observations.

\subsection{Impact of the mixed phase on hybrid star properties}

In order to investigate the possible correlation between the properties of the mixed phase and the global properties of neutron stars, we present the variation of the maximum mass ($M_{\mathrm{max}}$) as a function of the density extension of the mixed phase ($\Delta \rho$) in Fig.~\ref{fig:M_max_delta_rho}. The results are shown for variations of the quark matter parameters.

As shown in Tables~\ref{tab:mstar_ksat} and~\ref{tab:J_L}, the nuclear matter parameters have only a negligible impact on $M_{\mathrm{max}}$, and therefore the dominant effects arise from the quark sector.

In the case of the bag constant $B$, a non-monotonic behavior is observed for weak vector coupling, while for strong vector interaction $M_{\mathrm{max}}$ increases with $\Delta \rho$.
The variations in the vector coupling $G_V$ exhibit a strong positive correlation between $\Delta \rho$ and $M_{\mathrm{max}}$, indicating that enhanced repulsive interactions significantly stiffen the equation of state in the mixed-phase regime and increase the maximum supported mass.

\subsection{Summary of main results}

\textbf{Symmetric nuclear matter parameters:}
The mass--radius relation and tidal deformability are strongly sensitive to $m^*/m$, while $K_{\text{sat}}$ has only a weak effect; larger $G_V$ supports higher maximum masses consistent with observations.

\textbf{Symmetry energy and slope parameters:}
The symmetry energy parameters $J$ and $L$ mainly affect the hadronic branch by increasing stellar radii and tidal deformability, with minimal impact on the quark branch and maximum mass.

\textbf{Quark matter parameters:}
The bag constant $B_0^{1/4}$ and vector coupling $G_V$ govern the onset of quark matter and maximum mass, where larger $G_V$ yields heavier stars and smaller $G_V$ favors lower tidal deformability.

\textbf{Finite-temperature effects:}
Increasing temperature softens the equation of state, leading to larger radii, reduced maximum mass, and enhanced tidal deformability, partially counteracted by strong vector interaction.

\textbf{Impact of mixed phase:}
The maximum mass shows a strong positive correlation with the mixed-phase width $\Delta \rho$, dominated by quark matter parameters, particularly $G_V$.
\section{Summary and conclusions }
This study explores in detail the role of nuclear and quark parameters on the  thermodynamic behavior of the quark–hadron mixed phase within the Gibbs construction at finite temperature in the core of hybrid stars.
By varying nuclear saturation and symmetry energy parameters as well as the quark matter parameters, we have shown how
the structure and extent of the mixed phase respond to changes in both hadronic and
quark degrees of freedom.
The mixed-phase width is found to be particularly sensitive to the effective mass and
symmetry energy, whereas the incompressibility and symmetry energy slope exert a weaker
influence, especially at elevated temperatures.
Thermal effects play a central role in shaping the phase structure.
An increase in temperature leads to a reduction of the mixed-phase width and a softening
of the equation of state in the coexistence region, driven by Gibbs phase equilibrium
constraints.
We have examined the thermodynamic decomposition of speed of sound and the behavior of the trace anomaly and its derivative.
In this work a systematic study concerning the the structural properties of the mixed phase  has been carried out, and the main conclusions can be summarized as follows.
\begin{enumerate}
    \item The effective mass ratio $m^*/m$ strongly influences the mass, radius, and mixed-phase properties, with distinct behaviors for weak and strong vector interactions, while the vector coupling $G_V$ primarily controls the maximum mass and overall stiffness of the EOS. In contrast, the incompressibility $K_{\text{sat}}$ has a comparatively weak effect, mainly inducing a reordering of configurations after the onset of the quark–hadron transition.
    \item Increasing \(J\) lowers the hadron--quark transition mass in both cases, yielding a small radius change for \(G_V=0.2\) but a much larger radius jump (\(\sim 2\,\mathrm{km}\)) for \(G_V=0.02\).
    Increasing \(L_{\rm sym}\) shifts the hadron--quark transition to lower masses and enlarges the radius jump, with \(\Delta R\) growing rapidly (exceeding \(2\,\mathrm{km}\)) for weak repulsion (\(G_V=0.02\)) but remaining more moderate and controlled for strong repulsion (\(G_V=0.2\)).
   \item Variations in \(J\) and \(L\) mildly reduce \(M_{\max}\) and moderately affect radii, with NICER observations favoring lower \(L\) and remaining fully consistent with strong quark-matter repulsion (\(G_V=0.2\)).
    \item From an observational standpoint, lower vector coupling strengths favor compact stars with small radii and tidal deformabilities consistent with GW170817 and low-radius pulsars, while higher $G_V$ is required to explain massive ($\gtrsim 2,M_\odot$) pulsars.
    \item Variations in the bag constant shift the mass–radius relations between different NICER constraints, highlighting a tension between low-radius canonical pulsars and high-mass observations that can be alleviated by strong quark-matter repulsion.
    \item Finite temperature softens the EoS, shifting hybrid stars to larger radii, reducing the maximum mass, and enhancing \(\Delta R\) (up to \(\sim 4\!-\!5\) km), while strong vector repulsion (\(G_V=0.2\)) mitigates these effects.
    \item 
    Low-temperature configurations with weak repulsion satisfy GW170817, whereas strong vector repulsion maintains smaller deformabilities and consistency with massive pulsar constraints even at higher temperatures.

\end{enumerate}

\bibliography{hs_gc}

\end{document}